\documentclass{article}
\usepackage{spconf,amsmath,graphicx}
\usepackage{multirow}
\usepackage{cite}
\usepackage{booktabs}
\usepackage{bm}

\title{Convolution-based channel-frequency attention for TEXT-INDEPENDENT SPEAKER VERIFICATION}
\name{Jingyu Li, Yusheng Tian, Tan Lee}
\address{Department of Electronic Engineering, The Chinese University of Hong Kong, Hong Kong}
\begin{document}
\ninept
\maketitle
\begin{abstract}

Deep convolutional neural networks (CNNs) have been applied to extracting speaker embeddings with significant success in speaker verification. Incorporating the attention mechanism has shown to be effective in improving the model performance. This paper presents an efficient two-dimensional convolution-based attention module, namely \textbf{C2D-Att}. The interaction between the convolution channel and frequency is involved in the attention calculation by lightweight convolution layers. This requires only a small number of parameters. Fine-grained attention weights are produced to represent channel and frequency-specific information. The weights are imposed on the input features to improve the representation ability for speaker modeling. The \textbf{C2D-Att} is integrated into a modified version of ResNet for speaker embedding extraction. Experiments are conducted on VoxCeleb datasets. The results show that \textbf{C2D-Att} is effective in generating discriminative attention maps and outperforms other attention methods. The proposed model shows robust performance with different scales of model size and achieves state-of-the-art results.

\end{abstract}


\section{Introduction}
\label{sec:intro}

Speaker verification (SV) aims to determine whether the input speech utterance's speaker matches the claimed speaker\cite{campbell1997speaker,hansen2015speaker}. The present study is focused on text-independent SV, in which the input utterance has unrestricted content. A common SV system consists of a front-end embedding extraction model and a back-end score estimation method. The front-end model transforms the high-dimensional input speech into a compact vector, namely embedding, to represent speaker-specific characteristics. The back-end scoring method, e.g., cosine similarity and PLDA\cite{ioffe2006probabilistic}, measures the similarity between the embedding of input speech and the claimed speaker's embedding to give the decision of acceptance or rejection.

Deep neural network (DNN) based speaker embeddings such as X-vector and r-vector\cite{variani2014deep,snyder2018x,Nagrani17} showed superior performance to the factor analysis-based I-vector\cite{dehak2010front}. The input to the front-end model is usually spectral representations, e.g., Mel-scale filterbanks (Fbanks), as input. The input spectrogram represents information along the time and frequency dimensions. Convolutional neural networks (CNNs) were adopted in state-of-the-art speaker embedding networks.
1D-convolutional layers with dilation were used in the TDNN \cite{snyder2018x} and ECAPA\cite{desplanques2020ecapa} SV models, where frequency was treated as the channel dimension. Another line of work\cite{kwon2021ins,xie2019utterance,li2020text} treats the input spectrogram as an image with a single color channel, and uses 2D CNNs for feature extraction. As a result, the frequency dimension is retained in the extracted features. The front-end embedding extraction models discussed below are based on 2D CNNs.


The attention mechanism plays an important role in DNN-based embedding extraction models. The Squeeze-and-Excitation (SE) attention module was proposed in \cite{DBLP:journals/pami/HuSASW20} and applied successfully to the SV task\cite{zhou19_interspeech,desplanques2020ecapa}. The attention weights in SE are applied to the convolutional channel dimension. It does not exploit frequency-specific attention information, because features from different frequency regions within the same channel share the same weight. 
To better utilize frequency-specific information in the attention mechanism, frequency-wise SE was developed in \cite{thienpondt21_interspeech}. Channel-frequency interaction is involved in attention calculation by a multi-layer perceptron (MLP)\cite{DBLP:conf/icassp/LiuDLL22}. Discrete cosine transform (DCT) is utilized to re-scale the attention weights for different frequency and channel regions in \cite{sang2022multi}. Despite the effectiveness of such frequency-channel attention modules, the model size is significantly increased.

This paper proposes an efficient attention calculation module named \textbf{C2D-Att} to derive fine-grained channel-frequency attention weights for 2D CNN-based speaker embedding networks. The module consists of a pooling layer and two 2D convolutional layers. The pooling layer produces a channel-frequency plane by aggregating input features over the time dimension. Convolution operations are applied on the pooled channel-frequency plane to calculate the attention weights for different channel and frequency positions. Compared with the SE module, attention calculation in \textbf{C2D-Att} is more efficient in two aspects: $(1)$ it uses fewer parameters, $(2)$ channel-frequency interaction is involved, instead of calculation on one of the dimensions. 
ResNet\cite{he2016deep} is adopted as the network backbone structure in the experiments. Experimental results on the VoxCeleb datasets show that the proposed model significantly outperforms ResNet-SE and other state-of-the-art (SOTA) models.

The rest of the paper is organized as follows: Section \ref{sec:proposed_model} describes the attention mechanism of the proposed \textbf{C2D-Att} and the model structure. Sections \ref{sec:exp_setting} and \ref{sec:exp_result} give the experimental setting and results, respectively. Finally, we give a brief conclusion in Section \ref{sec:conclusion}.

\section{Related works}
\label{sec:related_works}

\subsection{Revisiting the SE attention module}
\label{ssec:SE}
Considering input feature $\bm{X} \in R^{C\times F \times T}$, where $C$ is the number of convolutional channels, $F$ and $T$ denote the lengths in the frequency and time dimensions, respectively. 
The first step in the SE module is global average pooling on the $F \times T$ plane as follows,

\begin{equation}
  s_c = \frac{1}{F \times T}\sum_{f=i}^{F}\sum_{t=j}^{T}\bm{X}(c,i,j)
  \label{eq:gap}
\end{equation}
The global information of each channel is aggregated in the vector $\bm{s}$ with length $C$. It is processed by two fully-connected layers (FC) and a non-linearity layer to capture the cross-channel interaction as,
\begin{equation}
  \bm{\omega} = \sigma(\textbf{W}_2(ReLU(\textbf{W}_1\bm{s})))
  \label{eq:se}
\end{equation}
where $\sigma(\cdot)$ is the sigmoid function, $\textbf{W}_2$ and $\textbf{W}_1$ are matrices with sizes $C \times d$ and $d \times C$, respectively. $d$ is a hyperparameter smaller than $C$ for reducing model complexity. The attention weight $\bm{\omega}\in R^C$ is applied on $\bm{X}$ to re-scale each channel by element-wise product. The SE module involves $2C\times d$ parameters.

\subsection{Attention in frequency dimension}
\label{ssec:freq_SE}
Fine-grained frequency information is crucial for speaker characteristics in speech data. The SE module averages over different frequencies to generate a global representation $\bm{s}$, in which all frequency regions share the same attention weight $\omega_c$ in one channel. This frequency-absent attention mechanism is not appropriate for representing and modeling speech data. \cite{thienpondt21_interspeech} proposes the frequency-wise SE ($fw$SE), where the dimension of $C$ in SE is replaced by $F$. The extracted attention weights are imposed on the frequency dimension for feature re-scaling. The number of parameters in frequency-wise SE equals $2F\times d$. 

In \cite{DBLP:conf/icassp/LiuDLL22}, both the frequency and channel dimensions of $\bm{X}$ are retained in the pooling process as follows,
\begin{equation}
  z_(c,f) = \frac{1}{T}\sum_{t=j}^{T}\bm{X}(c,f,j)
  \label{eq:gap_t}
\end{equation}
The channel-frequency plane $\bm{z} \in R^{C\times F}$ is then flattened into a 1D tensor and processed by an MLP layer for calculating the attention weights. This process leads to different attention weights on different frequency and channel regions. Meanwhile, it requires a large number of parameters. The parameters can be reduced by dividing features along the frequency dimension into $G$ groups, giving $2C\times F\times d/G$.

\begin{figure}[t]
  \centering
  \includegraphics[width=\linewidth]{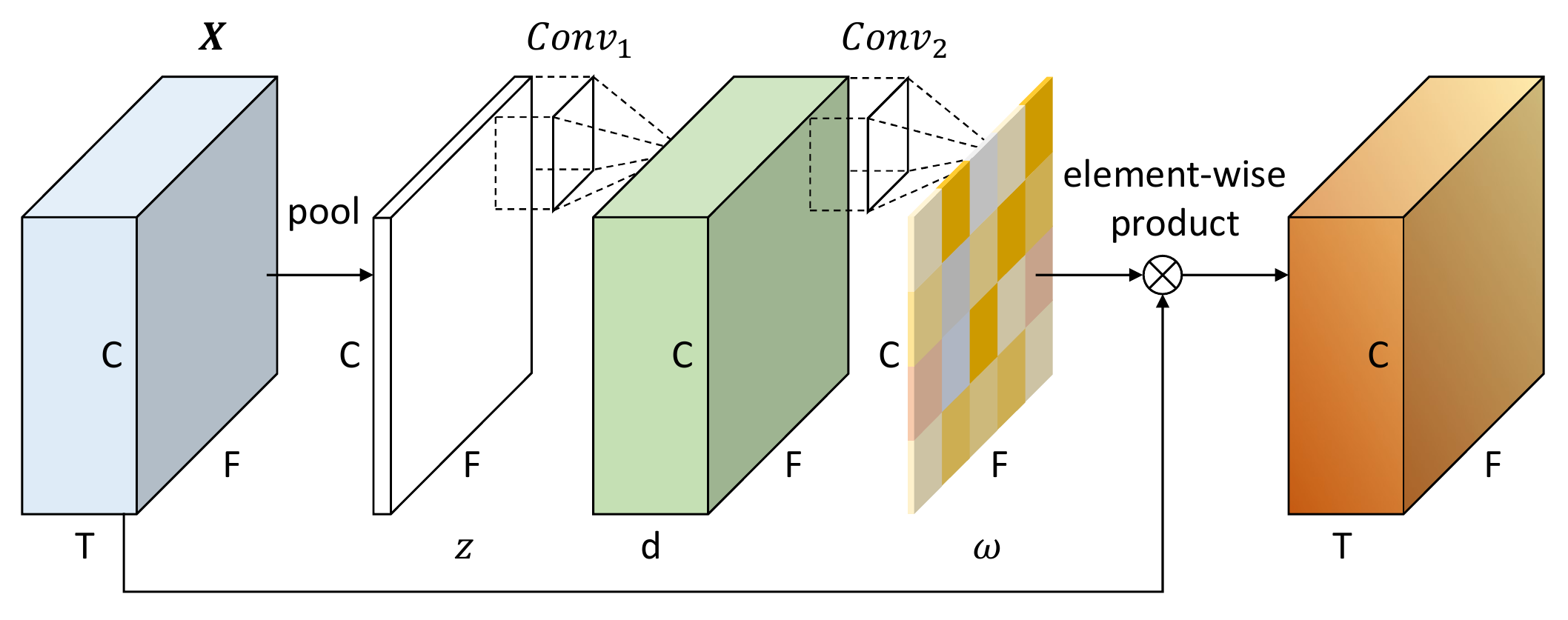}
  \caption{The network structure of the \textbf{C2D-Att} module. The non-linear layers and normalization layer are ignored. Best viewed in colors}
  \label{fig:c2d}
  \vspace{-2mm}
\end{figure}

\begin{table}
        \caption{The details of the model structure}
        \vspace{1mm}
        \label{tab:resnet_structure}
        \small{
        \centering
        \setlength\tabcolsep{9.5pt}
        \renewcommand{\arraystretch}{1.28}
        \begin{tabular}{lcc}
        \toprule 
        \textbf{Layer}  & \textbf{Structure} & \textbf{Output size} \\
        \hline
        \hline
        IN & --- & $F \times T$ \\
        \hline
        Conv1 & $7 \times 7, C$, stride $1$ & $C \times F \times T$ \\
        \hline
        Res1 & $\begin{bmatrix}
            3 \times 3, C\\
            3 \times 3, C\\
            C2D\textit{-}Att
            \end{bmatrix}$ $\times 3$, stride $1$ & $C \times F \times T$ \\
        \hline
        Res2 & $\begin{bmatrix}
            3 \times 3, 2C\\
            3 \times 3, 2C\\
            C2D\textit{-}Att
            \end{bmatrix}$ $\times 4$, stride $2$ & $2C \times F/2 \times T/2$ \\
        \hline
        Res3 & $\begin{bmatrix}
            3 \times 3, 4C\\
            3 \times 3, 4C\\
            C2D\textit{-}Att
            \end{bmatrix}$ $\times 6$, stride $2$ & $4C \times F/4 \times T/4$ \\
        \hline
        Res4 & $\begin{bmatrix}
            3 \times 3, 8C\\
            3 \times 3, 8C\\
            C2D\textit{-}Att
            \end{bmatrix}$ $\times 3$, stride $2$ & $8C \times F/8 \times T/8$ \\
        \hline
        Flatten & --- & $CF \times T/8$ \\
        \hline
        ASP & $\begin{bmatrix}
            CF \times 128\\
            128 \times CF \\
            \end{bmatrix}$ & $2CF$ \\
        \hline
        FC & $2CF \times E$ & $E$ \\
        \bottomrule 
        \end{tabular}
        \renewcommand{\arraystretch}{1}
    }
    \vspace{-2mm}
\end{table}

\section{Proposed model}
\label{sec:proposed_model}

Single-dimension attention does not exploit fine-grained information in the feature, and MLP-based channel-frequency attention calculation requires a large number of parameters. A lightweight channel-frequency convolution-based attention module is proposed in this paper. Attention weight calculation using convolution was proposed in \cite{DBLP:conf/cvpr/WangWZLZH20}, and better performance was obtained than the MLP-based module with significantly fewer parameters. Inspired by this, we propose to utilize convolution layers to compute the attention weights. In order to capture channel-frequency interaction on the plane $\bm{z}$ (from Eq.~\ref{eq:gap_t}), 2D convolution is adopted here,
\begin{equation}
  \bm{\omega} = \sigma(Conv_2(ReLU(BN(Conv_1(\bm{z})))))
  \label{eq:c2d_se}
\end{equation}
where $BN$ is a 2D batch normalization layer\cite{ioffe2015batch} and $Conv$ represents a 2D convolution layer. The 2D $Conv$ calculates the attention weights by aggregating the neighbor channels and frequencies of $\bm{z}$. Two $Conv$ layers are stacked to enhance the receptive field and learning ability of this module. $\bm{\omega} \in R^{C\times F}$ denotes the calculated attention weights on different frequency and channel positions. It is element-wise multiplied with $\bm{X}$. The structure is shown as in Fig.~\ref{fig:c2d}. Each $Conv$ has kernel size $k \times k$, and the middle channel size is denoted by $d$. Thus $2 \times k^2 \times d$ parameters are required in the \textbf{C2D-Att} module, excluding the parameters in the $BN$. The values of $k$ and $d$ used in our model are $3$ and $8$, respectively, giving only $144$ parameters. This module produces frequency and channel-specific attention weights, utilizing much fewer parameters than other methods discussed above.

The ResNet34\cite{he2016deep} is utilized as our baseline front-end model. Its model structure is modified to better fit the SV task. The input spectrogram is normalized on each frequency bin via an instance normalization (IN) layer\cite{DBLP:journals/corr/UlyanovVL16}. At the bottom of the model, the dimensions of channel and frequency are flattened and attentive statistics pooling (ASP)\cite{okabe2018attentive} is applied to the feature for time aggregation. An $E$-dimensional speaker embedding is extracted using the aggregated feature through a FC layer. $E=256$ is used in our experiments. \textbf{C2D-Att} is inserted at the end of each Res block but before the residual connection, which is the same as SE module\cite{DBLP:journals/pami/HuSASW20}. This model is named ResNet34-C2D in our experiments, and its details are given as in Table~\ref{tab:resnet_structure}.


In addition to average pooling (Eq.~\ref{eq:gap_t}), a second-order pooling method is investigated. Following \cite{wang2021revisiting}, we utilize the standard deviation ($std$) along the time dimension of $\bm{X}$ as the channel-frequency feature $\bm{z}$ and the performance gain is observed. This pooling method is denoted as $std$ pooling in the following.


\begin{table}[t]
  \caption{Performances of the models}
  \vspace{1mm}
  \label{tab:baseline_result}
  \centering
  \scalebox{0.99}{
  \begin{tabular}{lcccc}
    \toprule
    \multirow{2}{*}{\textbf{Model}} & \textbf{Params} & \multicolumn{3}{c}{\textbf{EER(\%)}} \\  
    \cline{3-5}
                            & \textbf{(M)}     &\textbf{Vox.O}       &\textbf{Vox.E}    &\textbf{Vox.H}   \\
    \midrule
    ResNet34                & 6.9              & 1.101               & 1.155            & 2.081           \\
    ResNet34-SE             & 6.98             & 0.957               & 1.198            & 2.202           \\
    ResNet34-$fw$SE         & 6.91             & 0.963               & 1.12             & 2.022           \\
    ResNet34-C2D            & 6.9              & \textbf{0.899}      & 1.11             & 2.044           \\
    \midrule
    ResNet34-SE$(std)$      & 6.98             & 0.979               & 1.158            & 2.088           \\
    ResNet34-$fw$SE$(std)$  & 6.91             & 0.989               & 1.156             & 2.089           \\
    ResNet34-C2D$(std)$     & 6.9              & \textbf{0.899}      & \textbf{1.03}    & \textbf{1.941}  \\
    \bottomrule
  \end{tabular}}
  \vspace{-2mm}
\end{table}

\begin{figure}[t]
  \centering
  \includegraphics[width=0.92\linewidth]{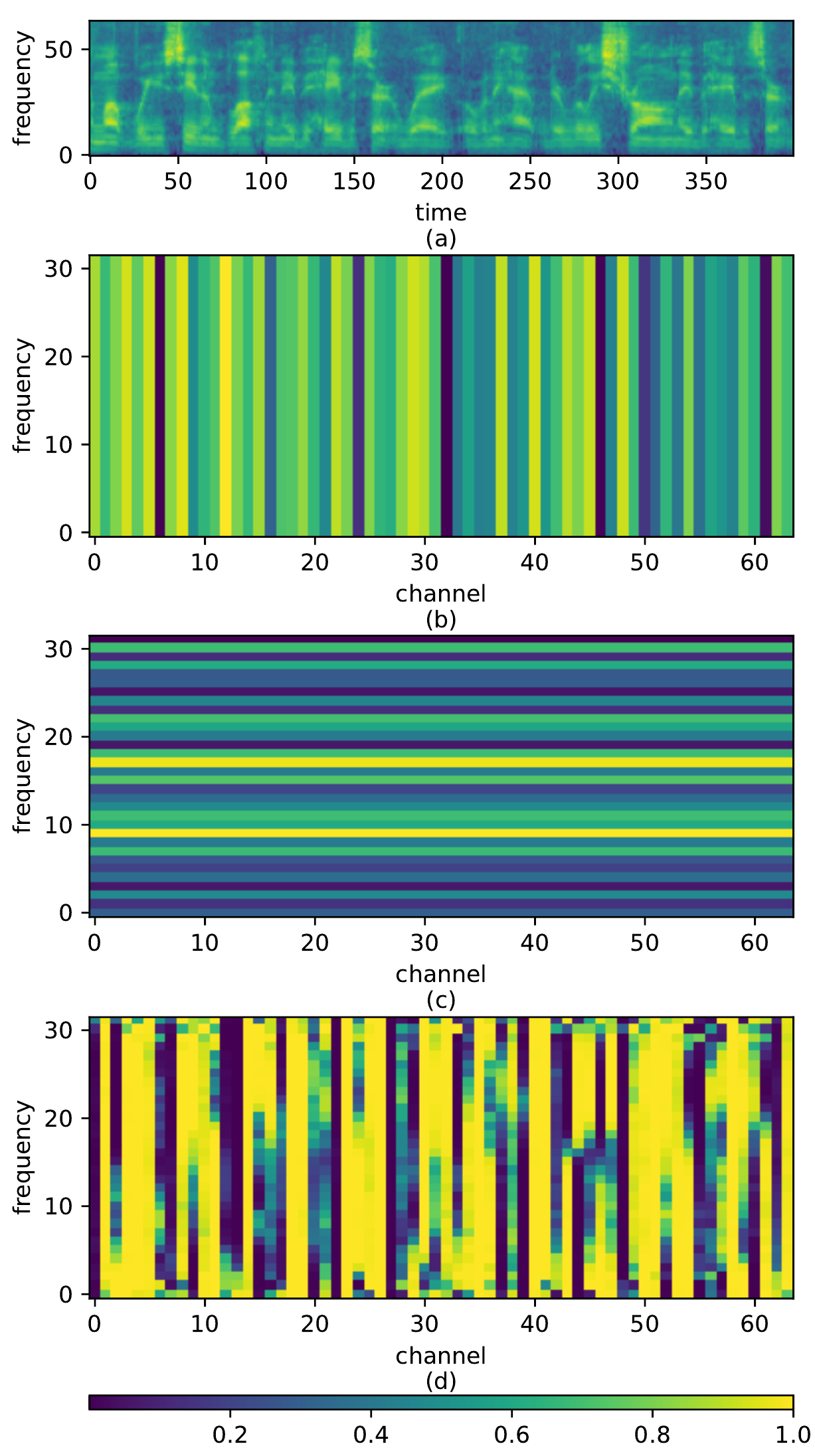}
  \caption{(a) $log$ Fbanks, and the attention weights given by (b) SE, (c) $fw$SE, (d) \textbf{C2D-Att}. Best viewed in colors}
  \label{fig:c2d_att}
  \vspace{-2mm}
\end{figure}

\section{Experimental settings}
\label{sec:exp_setting}

\subsection{Datasets}
\label{ssec:datasets}
VoxCeleb\cite{Nagrani17,Chung18b,Nagrani19} is a widely used dataset for text-independent SV research. In the present study, the development set (Dev) of VoxCeleb v2 (Vox.2) is adopted as the training set. It contains over $1$ million utterances from $5,994$ speakers. The model performance is evaluated on the original test set, the easy set and the hard set of VoxCeleb v1. They are denoted as Vox.O, Vox.E and Vox.H, respectively. Vox.E and Vox.H have more test pairs than Vox.O and are considered more challenging.

The input speeches are pre-emphasized with a first-order coefficient of $0.97$. $512$-point Fast Fourier transform (FFT) is applied with Hamming window of $25$ ms long and $10$ ms hop size. $64$-dimension $log$ Fbanks derived from the FFT spectrum are extracted as input features in our experiments.

\subsection{Data augmentation}
\label{ssec:augmentation}
Data augmentation is an effective technique for improving the robustness of SV systems\cite{cai2020fly,kwon2021ins}. Two augmentation methods are applied in this study, i.e., namely background sound addition and audio reverberation. The background sound is randomly sampled from MUSAN\cite{snyder2015musan}, which consists of three types of interfering sounds: babble, noise, and music. Audio reverberation is implemented by convolving input speech with simulated room impulse responses randomly selected from the dataset \cite{ko2017study}.

\subsection{Training procedure}
\label{ssec:training}
In each training step, a 2-second segment is randomly cropped from each input utterance. Data augmentation is applied as described above. Speaker embeddings are extracted from the speech segments using the front-end model. The Additive Angular Margin softmax (AAM-softmax)\cite{deng2019arcface} function is used as the identification loss. The margin and scale of AAM-softmax are set to be $0.2$ and $30$, respectively. The loss is optimized by the Adam optimizer\cite{DBLP:journals/corr/KingmaB14} with a weight decay of $2e\textit{-}5$.
Step decay procedure with warming up is applied to adjust the learning rate. The learning rate is initialized as $0$ and increased linearly to $0.001$ in $20,000$ steps.  
It is decreased by a ratio of $0.1$ at the $20th$ and $32nd$ epoch. The training process is stopped after $40$ epochs.

\subsection{Evaluation protocol}
\label{ssec:evaluation}
For performance evaluation, each test utterance is divided into $4$-second long segments, with $1$-second overlap between neighboring segments. Speaker embeddings are extracted from all segments. Cosine similarities between all pairs of test embeddings and enrollment embeddings are averaged to give the overall score for SV decision. The scores are calibrated by the adaptive s-norm (AS-norm)\cite{matejka2017analysis}. $10,000$ utterances are randomly selected from the Dev of Vox.2 as the imposter cohort for the AS-norm, and the imposters with top $1,000$ scores are chosen for our experiments.

\section{Results}
\label{sec:exp_result}

\begin{table*}[t]
  \caption{The EER($\%$) and minDCF of different models. $P_{target}=0.01$, $C_{miss}=C_{fa}=1$ are used to calculate the minDCF metric}
  \vspace{1mm}
  \label{tab:sota_result}
  \setlength\tabcolsep{3.5pt}
  \renewcommand{\arraystretch}{0.95}
  \centering
  \begin{tabular*}{\textwidth}{c c c @{\extracolsep{\fill}} cc cc cc}
    \toprule
    \multirow{2}{*}{\textbf{Model}} & \multirow{2}{*}{\textbf{Params}} &
    \multirow{2}{*}{\textbf{Input}} &
    \multicolumn{2}{c}{\textbf{Vox.O}} & \multicolumn{2}{c}{\textbf{Vox.E}} & \multicolumn{2}{c}{\textbf{Vox.H}}  \\
     \cmidrule{4-9}
    & \textbf{(M)} & \textbf{feature} & EER & minDCF &  EER & minDCF &  EER &minDCF   \\
    \midrule
    \midrule
    E-TDNN\cite{desplanques2020ecapa} &  6.8 & 80-dim MFCC & $1.49$ & $0.1604$ & $1.61$ & $0.1712$ & $2.69$ & $0.2419$   \\
    \midrule
    E-TDNN(large)\cite{desplanques2020ecapa} & 20.4 & 80-dim MFCC& $1.26$ & $0.1399$ & $1.37$ & $0.1487$ & $2.35$ & $0.2153$  \\
    \midrule
    ECAPA-TDNN-512\cite{desplanques2020ecapa} & 6.2 &80-dim MFCC & $1.01$ & $0.1274$ & $1.24$ & $0.1418$ & $2.32$ & $0.2181$ \\
    \midrule
    ECAPA-TDNN-1024\cite{desplanques2020ecapa} & 14.7 &80-dim MFCC & $0.87$ & $0.1066$ & $1.12$ & $0.1318$ & $2.12$ & $0.2101$  \\
    \midrule
    ECAPA CNN-TDNN-512\cite{DBLP:conf/icassp/LiuDLL22} & 7.66 & 80-dim Fbanks & $0.92$ & $0.0921$ & $1.111$ & $0.1154$ & $2.141$ & $0.2036$   \\
    \midrule
    MFA-TDNN-512\cite{DBLP:conf/icassp/LiuDLL22} & 7.32 & 80-dim Fbanks & $0.856$ & $0.0923$ & $1.083$ & $0.1175$ & $2.049$ & $0.1897$  \\
    \midrule
    ResNet34-SFSC\cite{sang2022multi} & 8.0 & 64-dim Fbanks & $0.97$ & - & $1.2$ & - & $2.29$ & - \\
    \midrule
    ResNet34-MFSC\cite{sang2022multi}  & 8.0 & 64-dim Fbanks & $0.91$ & - & $1.21$ & - & $2.3$ & - \\
    \midrule
    ResNet34-C2D-25 & 4.49 & 64-dim Fbanks   & $0.995$ & $0.10428$ & $1.112$ & $0.1294$ & $2.113$ & $0.209$  \\
    \midrule
    ResNet34-C2D-32 & 6.9 & 64-dim Fbanks   & $0.899$  & $0.1034$  & $1.03$  & $0.1247$ & $1.941$ & $0.1917$ \\
    \midrule
    ResNet34-C2D-32 & 7.3 & 80-dim Fbanks   & $0.808$  & $0.0918$  & $1.031$ & $0.1195$ & $1.921$ & $0.1996$ \\
    \midrule
    ResNet34-C2D-40 & 10.29 & 64-dim Fbanks & $\bm{0.75}$ & $\bm{0.081}$ & $0.998$ & $0.1128$ & $1.877$ & $0.186$  \\
    \midrule
    ResNet52-C2D-32 & 10.34 & 64-dim Fbanks & $0.771$ & $0.1071$ & $\bm{0.939}$ & $\bm{0.111}$ &                                                                 $\bm{1.816}$ & $\bm{0.1796}$  \\
    \bottomrule
 \end{tabular*} 
\end{table*}

\subsection{Comparison with the baselines}
\label{ssec:baseline}
The ResNet34 and its variants are used as the baselines with $C=32$ (Table~\ref{tab:resnet_structure}). The equal error rates (EER) of different models are reported as in Table~\ref{tab:baseline_result}. The ResNet-SE outperforms the ResNet only on the Vox.O set, meaning that channel-wise attention has little impact on model performance. Frequency-wise SE ($fw$SE) involves frequency positional encoding in the model. It is considered to be a strong baseline and shows a lower EER than the ResNet on all test sets. This suggests that frequency information is crucial for extracting representative speaker features. The proposed ResNet-C2D demonstrates significant performance gain over the baselines. Its EER is lower than those of the SE and $fw$SE, except for Vox.H on $fw$SE. With 2D convolution, fewer parameters are required in the \textbf{C2D-Att} than SE and $fw$SE. 

In addition, the average pooling in the attention modules is replaced by $std$ pooling. Interestingly, better results can be achieved in SE and \textbf{C2D-Att}, while the performance of $fw$SE declines. \textbf{C2D-Att} achieves the best performance among all models. This indicates that the second-order pooling improves the robust aggregation ability in the proposed attention module. $std$ is therefore applied in the pooling layers of ResNet-C2D in the following experiments.

\subsection{Comparison with SOTA models}
\label{ssec:sota}

ResNet-C2D is compared with several state-of-the-art (SOTA) models. The results are given as in Table~\ref{tab:sota_result}. In addition to EER, minimum Detection Cost Function (minDCF) is reported.
The proposed ResNet34-C2D-32 shows better performance on Vox.E and Vox.H than ECAPA-TDNN-1024 and E-TDNN(large), with half the number of model parameters. ECAPA CNN-TDNN and MFA-TDNN are two variants of ECAPA-TDNN with similar model sizes. Lower EERs can be achieved by ResNet34-C2D-32 than these two stronger models.

The model scalability is evaluated by using different input sizes ($64$ vs. $80$), channel widths ($25$, $32$, $40$) or the model depths ($34$ vs. $52$), respectively. The channel width is denoted as the ending digits in the model names. The structure of ResNet52-C2D-32 is built by increasing the number of Res blocks in Table~\ref{tab:resnet_structure} to $\{5$, $6$, $9$, $5\}$. ResNet34-C2D-25 outperforms ECAPA-TDNN-512 on all three test sets, with $30\%$ fewer parameters. The performance of the network is improved as the model complexity increases, indicating the robustness of the network with different model complexities. ResNet34-C2D-40 achieves the best performance on Vox.O, and ResNet52-C2D-32 gives the best results on Vox.E and Vox.H. 

\subsection{Visualization}
\label{ssec:visual}
 A speech utterance is selected randomly from the dataset. Its $log$ Fbanks spectrogram is shown as in Fig.~\ref{fig:c2d_att}(a). The attention maps are calculated by three different methods and plotted in (b,c,d) of Fig.~\ref{fig:c2d_att}. They are derived at the last block of Res2 with a size of $64 \times 32$. The x-axis denotes the channel dimension, and the y-axis represents frequency. In (b) and (c), different weights are imposed on only the channel or frequency dimension. In (d), it can be observed that the weights for different channel and frequency positions are not identical. Therefore the channel-frequency attention mechanism has a stronger representation ability than the only channel-wise or frequency-wise attention.

\section{Conclusions}
\label{sec:conclusion}
A novel 2D convolution-based attention mechanism, called \textbf{C2D-Att}, is proposed for speaker verification. The proposed method captures local channel-frequency interaction to determine the attention weights for different channel-frequency positions. With the use of convolution, fewer model parameters are required in \textbf{C2D-Att}, making the calculation more efficient. \textbf{C2D-Att} is combined with a modified version of ResNet as the speaker embedding extraction model. The proposed model demonstrates significant performance improvement on VoxCeleb in our experiments, compared with other attention methods, e.g., SE and $fw$SE. This shows that the channel-frequency attention is suitable for capturing useful speaker information. The proposed model is evaluated under different model scales and outperforms several SOTA models using a similar or fewer number of parameters.

\section{Acknowledgements}
\label{sec:acknowledgements}
The first two authors are supported by the Hong Kong PhD Fellowship Scheme.

\bibliographystyle{IEEEbib}
\bibliography{refs}

\end{document}